\documentclass[pdflatex,sn-standardnature]{sn-jnl}
\usepackage{float}
\usepackage{amsmath,amsthm}

\theoremstyle{thmstyleone}%

\theoremstyle{thmstyletwo}%

\theoremstyle{thmstylethree}%

\begin{document}

\title{Wafer-Scale Squeezed-Light Chips}

\author[1]{Shuai Liu}
\author[1]{Kailu Zhou}
\author[1]{Yuheng Zhang}
\author[1]{Abdulkarim Hariri}
\author[1]{Nicholas Reynolds}
\author[2]{Bo-Han Wu}
\author*[1]{Zheshen Zhang}\email{zszh@umich.edu}

\affil[1]{\orgdiv{Department of Electrical Engineering and Computer Science}, \orgname{University of Michigan}, \orgaddress{\city{Ann Arbor}, \postcode{48109}, \state{Michigan}, \country{USA}}}
\affil[2]{\orgdiv{Department of Electrical and Computer Engineering}, \orgname{University of Hawai'i at Mānoa}, \city{Honolulu}, \postcode{96822}, \state{Hawaii}, \country{USA}}

\maketitle

\abstract{}

Squeezed-light generation in photonic integrated circuits (PICs) is essential for scalable continuous-variable (CV) quantum information processing. By suppressing quantum fluctuations below the shot-noise limit, squeezed states enable quantum-enhanced sensing and serve as a standard resource for CV quantum information processing. While chip-level squeezed-light sources have been demonstrated, extending this capability to the wafer level with reproducible strong squeezing to bolster large-scale quantum-enhanced sensing and information processing has been hindered by squeezed light's extreme susceptibility to device imperfections. Here, we report wafer-scale fabrication, generation, and characterization of two-mode squeezed-vacuum states on a fully complementary metal-oxide-semiconductor (CMOS)-compatible silicon nitride (Si$_3$N$_4$) PIC platform. Across a 4-inch wafer, 8 dies yield 2.9-3.1 dB directly measured quadrature squeezing with $< 0.2$ dB variation, demonstrating excellent uniformity. This performance is enabled by co-integrating ultralow-loss, strongly overcoupled high-$Q$ microresonators, cascaded pump-rejection filters, and low-loss inverse-tapered edge couplers. The measurements agree with a first-principles theoretical model parameterized solely by independently extracted device parameters and experimental settings. The measured squeezing level can be further improved by enhancing the efficiencies of off-chip detection and chip-to-fiber coupling. These results establish a reproducible, wafer-scale route to nonclassical-light generation in integrated photonics and lay the groundwork for scalable CV processors, multiplexed entanglement sources, and quantum-enhanced sensing.


\section{Introduction}\label{sec:introduction}
Quantum science and technology are emerging as transformative drivers of innovation across advanced computing \cite{arute2019quantum, madsen2022quantum}, sensing \cite{pirandola2018advances, degen2017quantum}, and secure communication \cite{gisin2007quantum}, promising capabilities beyond the reach of classical systems. At the heart of many of these applications lies the ability to generate, manipulate, and detect non-classical states of light. Among them, squeezed light---in which quantum fluctuations in one quadrature are suppressed below the shot-noise limit (SNL) \cite{walls1983squeezed}---offers intrinsic quantum advantages in precision measurements and serves as a key resource for CV quantum information processing protocols \cite{villar2005generation, lawrie2019quantum, o2009photonic, masada2015continuous}. Since its first experimental demonstration in 1985 \cite{slusher1985observation}, squeezed light has been extensively explored in bulk optical platforms, achieving record shot-noise suppression exceeding 15 dB \cite{vahlbruch2008observation, schonbeck201713, vahlbruch2016detection}. Such table-top systems have enabled landmark achievements, from enhancing the sensitivity of gravitational-wave detectors (e.g., LIGO) \cite{aasi2013enhanced, ganapathy2023broadband} to advancing quantum imaging of biological structures \cite{casacio2021quantum}, dual-comb spectroscopy for gas sensing \cite{herman2025squeezed, hariri2024entangled}, generation of large-scale cluster states \cite{asavanant2019generation, larsen2019deterministic}, and distributed entangled sensor networks \cite{xia2020demonstration, guo2020distributed,xia2023entanglement,xia2021quantum}.

Translating these laboratory demonstrations into deployable quantum systems requires platforms that maintain high performance while battling scalability, stability, robust integration, and cost-effectiveness challenges \cite{moody20222022}. Photonic integrated circuits fabricated in CMOS-compatible processes provide a compelling route: they offer phase-stable operations, environmental robustness, and mature process design kits (PDKs) with wafer-scale fabrication control. Recent advances have realized on-chip CV cluster states \cite{wang2025large, jia2025continuous}, broadband homodyne detection integrated with CMOS electronics \cite{tasker2021silicon}, and antenna arrays for squeezed light \cite{gurses2025chip}. Across these applications, preserving strong squeezing is a pivotal requirement to achieve a quantum advantage---benchmarks for the amount of needed squeezing commonly cited in the field include at least 3 dB for enabling entanglement swapping \cite{van1999unconditional, masada2015continuous}, 4.5 dB for two-dimensional cluster-state generation \cite{asavanant2019generation}, and 10 dB for fault-tolerant CV quantum computing \cite{fukui2018high,fukui2023high}. However, squeezed light is extreme vulnerable to loss and noise stemming from practical imperfections due to, e.g., propagation loss incurred along the optical path, phase-locking instability, and the inefficiencies of optical components such as filters needed to reject the pump.

Over the past decade, chip-scale squeezed-light sources have been investigated across several material platforms---including thin-film lithium niobate (TFLN) \cite{chen2022ultra, park2024single, arge2024demonstration, shi2025squeezed}, Si$_3$N$_4$ \cite{dutt2015chip, vaidya2020broadband, zhao2020near, zhang2021squeezed, shen2025strong, ulanov2025quadrature}, and silica (SiO$_2$) \cite{yang2021squeezed, wang2025large}---each offering distinct trade-offs in nonlinearity, loss, cost, and process maturity. To deliver reproducible, high-performance squeezing at scale, fully CMOS-compatible PICs are especially attractive: they leverage established PDKs and foundry workflows while enabling co-integration of the classical, linear photonic units, such as filters, phase shifters, interferometers, and low-loss delay lines, to complement the quantum components, positioning Si$_3$N$_4$ as a leading candidate to meet these requirements \cite{xiang2022silicon, shekhar2024roadmapping}. A vital, albeit persistently challenging, step toward manufacturable quantum PICs has been migrating chip-level demonstrations to a wafer-scale production while preserving strong squeezing, as even modest excess loss and noise can rapidly erode the achievable quantum advantages in different CV quantum sensing and information processing tasks. In particular, achieving $>$ 3 dB of measured squeezing is recognized as a first critical milestone, marking the threshold for bona fide quantum advantages in CV quantum teleportation~\cite{grosshans2001quantum} and entanglement swapping~\cite{van1999unconditional} and deemed a transition from marginal noise reduction to operationally useful nonclassical resources for quantum-enhanced tasks. 

In this work, we address this challenge by demonstrating wafer-scale fabrication and generation of two-mode squeezed vacuum (TMSV) states on a fully CMOS-compatible Si$_3$N$_4$ PIC platform. We achieve $> 3$ dB of directly measured quadrature squeezing uniformly across a 4-inch wafer, with device-to-device variation $< 0.2$ dB (2.9-3.1 dB). This reproducible performance is enabled by a single subtractive fabrication process that co-integrates, for the first time, ultralow-loss ($Q_{\rm i}>10^7$), dispersion-engineered, strongly overcoupled microresonators (escape efficiency $> 91\%$ near 1560~nm), high-extinction pump-rejection filters (each $>\!30$~dB), and low-loss inverse-tapered edge couplers (chip--fiber coupling $>\!75\%$) on the same chip. By reducing insertion loss and providing stable thermo-optic control, this system-level integration converts squeezed-light generation from a device-level demonstration into a wafer-scale, manufacturable resource for CV quantum photonics. Together with recent progress in system-level electro-optical co-packaging, our results establish a new benchmark for reproducible, wafer-scale nonclassical-light generation in integrated photonics and point toward high-volume CMOS manufacturing of quantum processing circuits.

\section{Design of quantum photonic integrated circuits}

\begin{figure}[bth!]
\centering
\includegraphics[width=1\textwidth]{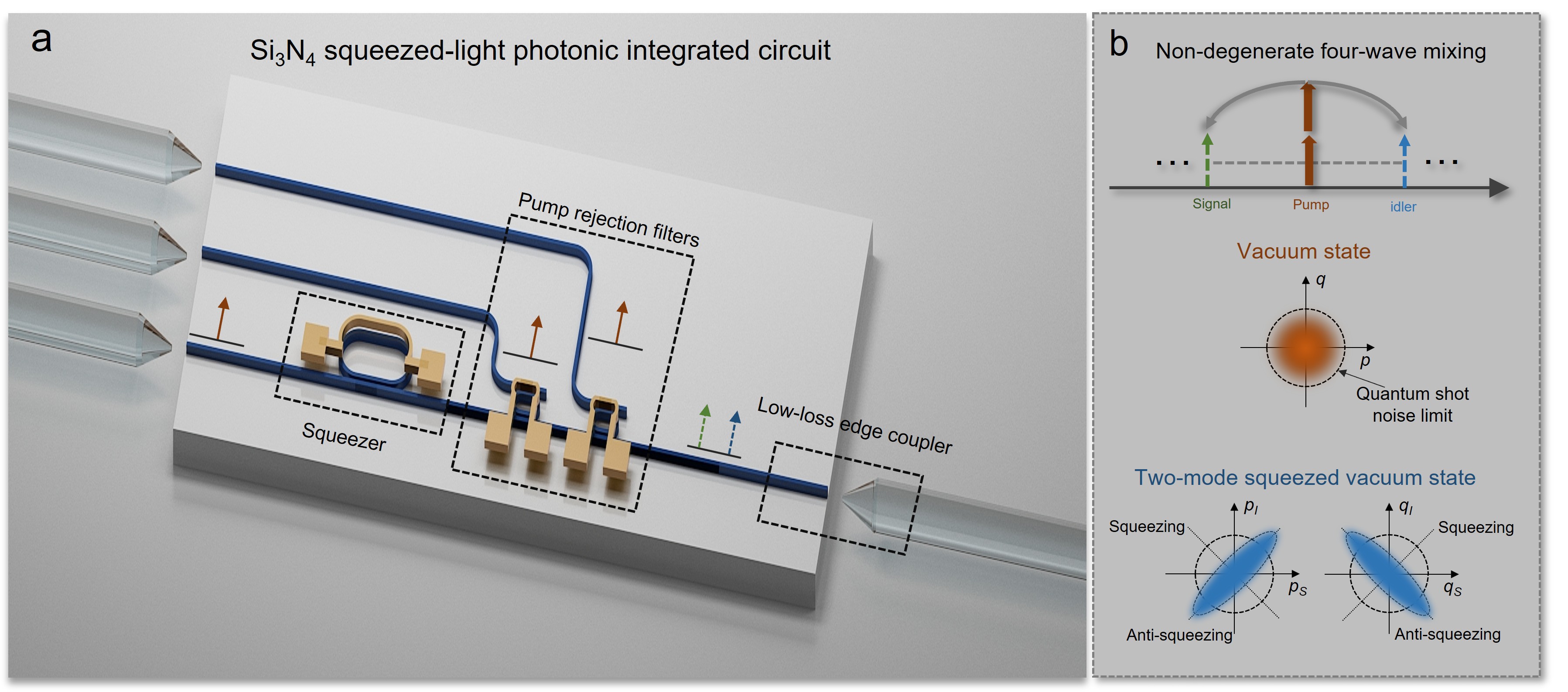}
\caption{ (a) Schematic of the Si$_3$N$_4$ PIC for TMSV generation. The circuit integrates a high-$Q$ microring resonator for squeezed-light generation via non-degenerate FWM, two cascaded add-drop filters with high extinction ratios for residual pump suppression, and an inverse-tapered waveguide edge coupler for low-loss fiber collection and off-chip characterization. (b) Conceptual illustration of the non-degenerate FWM process in the microring resonator: two pump photons are annihilated to generate quantum correlated signal and idler photons in adjacent resonances. The output signal and idler modes are in a TMSV state that exhibits correlated amplitude quadratures $q_S$ and $q_I$ and anti-correlated phase quadratures $p_S$ and $p_I$.}\label{fig:diagram}
\end{figure}

Figure \ref{fig:diagram}a illustrates the schematic of the device architecture for on-chip squeezed-light generation. A continuous-wave (CW) pump laser, operated below the parametric oscillation threshold, is coupled into a strongly over-coupled Si$_3$N$_4$ microring resonator. Inside the cavity, the enhanced circulating pump field drives non-degenerate four-wave mixing (FWM) via the $\chi^{(3)}$ nonlinearity of the Si$_3$N$_4$ waveguide, in which two pump photons are annihilated and converted into a pair of correlated signal and idler photons occupying adjacent resonances, as illustrated in Fig.~\ref{fig:diagram}b. The emitted signal-idler mode pairs form a TMSV state, exhibiting quadrature noise suppression of $\hat{p}_S+\hat{p}_I$ and $\hat{q}_S-\hat{q}_I$ below the shot-noise limit (SNL)—a defining signature of squeezing—while the conjugate quadratures $\hat{p}_S-\hat{p}_I$ and $\hat{q}_S+\hat{q}_I$ display increased noise fluctuation, i.e., anti-squeezing.

Strong squeezing is essential for CV quantum technologies, as the squeezing level directly determines the degree of quantum advantage attainable in sensing, communication, and information processing. The maximum achievable squeezing extracted from the resonator is governed by the escape efficiency \cite{dutt2015chip, yang2021squeezed}, defined as $\eta = 1 - Q_{\rm L}/Q_{\rm i}$, with $1/Q_{\rm L} = 1/Q_{\rm i} + 1/Q_{\rm c}$, where $Q_{\rm i}$, $Q_{\rm c}$, and $Q_{\rm L}$ are the intrinsic, coupling, and loaded quality factors. In principle, high $Q_{\rm i}$---typically limited by nanofabrication imperfections and material absorption---together with a moderately low $Q_{\rm L}$ is desirable for generating strong squeezing. Lowering $Q_{\rm L}$ increases the $\eta$ and thus the potential squeezing level, but it also leads to undesired effects such as a higher parametric oscillation threshold and increased susceptibility to photo-refractive noise from stronger pump power. In practice, the measured squeezing is limited by all losses incurred between the squeezing chip and the off-chip photodetectors, which constitute the dominant imperfection that caps the amount of measured squeezing level in our experiment. For $\chi^{(3)}$-based nanophotonics squeezed-light sources, an additional challenge arises because the pump laser lies spectrally close to the squeezed modes, making efficient pump rejection with minimal insertion loss a formidable task. To address this, we employ two integrated add-drop filters with large free spectral range (FSR) placed after the microring resonator that generates squeezed light, namely, the squeezer. These filters provide high extinction ratios, negligible insertion loss, and broadband resonance tunability, thereby efficiently suppressing residual pump light while minimizing system loss. Thermo-optic microheaters are integrated on both the squeezer and the filters for resonance tuning and long-term stabilization. The filtered squeezed light is then collected through an inverse-tapered waveguide coupler, interfered with a bi-tone local oscillator (LO), and characterized at a balanced homodyne detector (BHD) to retrieve the noise spectra across different quadratures.

\section{\label{sec:results} Results}

\subsection{Device fabrication and testing}

\begin{figure}[bth!]
\centering
\includegraphics[width=1\textwidth]{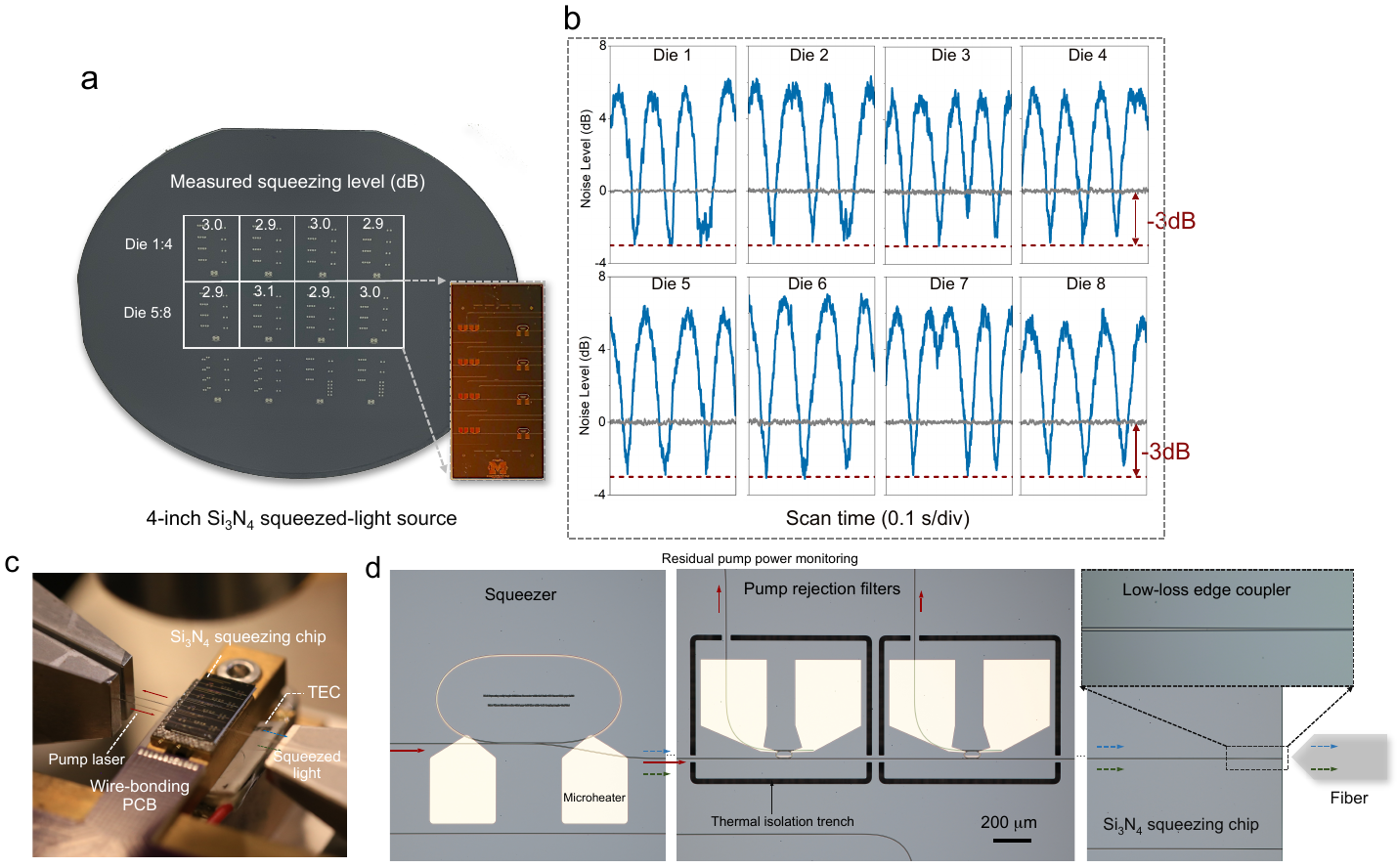}
\caption{(a) Optical image of a fabricated 4-inch Si$_3$N$_4$ quantum photonic wafer containing 8 dies of integrated squeezed-light sources. The directly measured maximum squeezing level of each die is labeled at its position, while four additional dies are reserved for other projects. A zoom-in image of one die is displayed at the side. Across the whole wafer, all dies exhibit measured squeezing levels close to 3 dB, ranging from 2.9 dB to 3.1 dB, with a variation of less than 0.2 dB. (b) Measured quadrature noise variances (blue trace) relative to the calibrated shot-noise level (gray trace) for the 8 dies. The LO phase is scanned to access both squeezed and anti-squeezed quadratures. (c) Photograph of a Si$_3$N$_4$ squeezing chip wire-bonded to a PCB for thermo-optic resonance detuning and stabilization. The chip is mounted on a temperature-stabilized stage with a PDH-controlled thermoelectric cooler and coupled with input and output light via lensed fibers. (d) Optical micrographs of the core functional elements of the Si$_3$N$_4$ squeezing chip: a racetrack microring squeezer for TMSV generation, add-drop pump-rejection filters to suppress residual pump light, and inverse-tapered waveguide couplers for efficient squeezed-light collection into optical fibers. }\label{fig:wafer}
\end{figure}

We fabricate the designed squeezing chips on a 4-inch, 800-nm-thick Si$_3$N$_4$-on-SiO$_2$-on-Si wafer (see Methods). Each wafer contains eight dies for TMSV generation, with each die integrating four squeezed-light photonic circuits implementing different pump-rejection filter parameters, while the remaining four dies are reserved for other projects. To benchmark wafer-scale squeezing performance, we generate and characterize TMSV states across all eight dies following a similar procedure as previous demonstrations (Ref.~\cite{yang2021squeezed} and Methods). In our experiment, a CW laser boosted by an erbium-doped fiber amplifier (EDFA) is used as the pump, with amplified spontaneous emission (ASE) noise suppressed by a Pound–Drever–Hall (PDH)-locked, 2 MHz-linewidth free-space mode cleaner. The squeezer's and filters' resonances are initially red-detuned and then gradually tuned toward the pump wavelength by reducing the microheater temperature, thereby stabilizing resonance aligned with the pump. The generated TMSV signal is collected at the through port. A phase-coherent LO is derived from the same CW pump by driving a fiber electro-optic modulator (EOM) to generate an EO comb followed by a programmable filter used to select bi-tone comb teeth frequency-matched to the two TMSV modes. Finally, the squeezed light and the LO are interfered on a 50:50 beam splitter and detected with a homemade BHD to measure the quadrature noise variances. 

Figure \ref{fig:wafer}a shows a photograph of the fabricated 4-inch Si$_3$N$_4$ squeezed-light wafer, with insets labeling the directly measured maximum squeezing level of each die. The corresponding quadrature noise-variance spectra, normalized to the calibrated shot-noise level, are shown in Fig.~\ref{fig:wafer}b. Four dies exceed or reach 3.0 dB of squeezing, with the highest level measured at 3.1 dB and the lowest at 2.9 dB. Across the wafer, the squeezing levels range from 2.9 to 3.1 dB, yielding an average of $2.96 \pm 0.2$ dB, demonstrating excellent wafer-scale reproducibility and uniformity.

Fig.~\ref{fig:wafer}c shows a photograph of a Si$_3$N$_4$ squeezing die wire-bonded to a customized printed circuit board (PCB) for thermo-optic resonance tuning and locking. Low-loss edge coupling is achieved using lensed fibers and inverse-tapered waveguides. Optical micrographs of the core functional elements of the Si$_3$N$_4$ squeezing chip are presented in Fig.~\ref{fig:wafer}d. The on-chip squeezer consists of a racetrack microring resonator composed of two modified Euler bends connected by straight sections. The resonator is side-coupled to a bus waveguide of identical cross-sectional geometry, which ensures inherent broadband phase matching and excitation of the low-loss, fundamental waveguide modes. A trade-off between minimizing intrinsic scattering loss (high $Q\mathrm{i}$), achieving strong bus-to-resonator coupling (low $Q_\mathrm{c}$ and $Q_\mathrm{L}$), maintaining near-zero anomalous dispersion, and suppressing mode coupling is realized with a waveguide cross section of 2600 nm $\times$ 800 nm and optimized Euler curvature. The total cavity length is approximately 2425 $\mu$m, corresponding to an FSR of about 59.3 GHz. To reject the strong residual pump, the output is passed through two cascaded add-drop microring filters, each of which employs a 1500 nm $\times$ 800 nm waveguide cross section with optimized Euler bends. The two bus-to-resonator couplings are optimized such that  the coupling loss dominates the intrinsic loss, placing the filters near the critical-coupling regime and rendering the extinction ratio robust against nanofabrication-induced variations or input-power–dependent material absorption. Their FSR ($\sim$603 GHz) is engineered to minimize the impact on the squeezing modes. The drop ports provide power-monitoring signals for active resonance locking via thermo-optic detuning. Deep thermal isolation trenches are introduced around the filters to suppress thermal crosstalk with the squeezer. Finally, the filtered TMSV light is edge-coupled out through a 300 $\mu$m-long inverse-tapered waveguide with a tip dimension of 220 nm $\times$ 800 nm, exhibiting good mode matching with a 5 $\mu$m-diameter lensed fiber to achieve low coupling loss. 

\begin{figure}[bth!]
\centering
\includegraphics[width=1\textwidth]{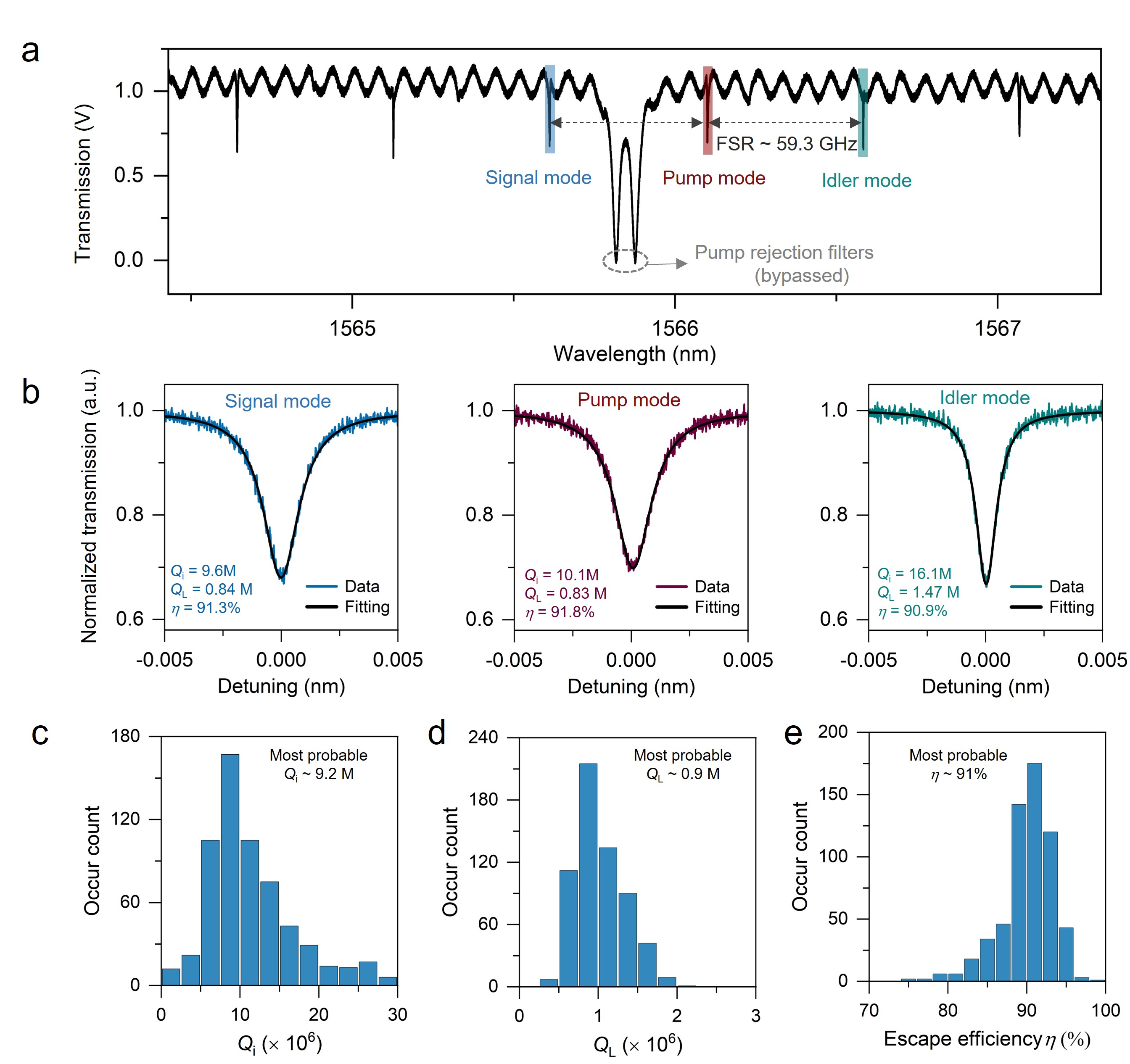}
\caption{Optical characterization of the fabricated Si$_3$N$_4$ squeezing chips. (a) A typical measured transmission spectrum of a Si$_3$N$_4$ squeezing chip, showing resonances due to a strongly over-coupled squeezer and two cascaded pump rejection filters. To measure the linewidths of the resonant modes, the filter resonances are detuned to bypass the squeezing modes, whereas both the squeezer pump mode and the filter resonances are tuned into alignment with the pump laser during squeezing experiments. (b) Normalized transmission spectra of the resonances selected as pump, signal, and idler modes for squeezed-light generation. The $Q_\mathrm{i}$ and $Q_\mathrm{L}$ factors are extracted by Lorentzian fitting, showing large escape efficiencies $\eta > 90\%$ while maintaining high $Q_\mathrm{L} > 8\times10^5$. (c)-(e) Statistical distributions of $Q_\mathrm{i}$, $Q_\mathrm{L}$, and $\eta$ for squeezer resonances in the 1550–1580 nm range, corroborating wafer-scale uniformity of the squeezing devices.}\label{fig:classical}
\end{figure}

We test the classical performance of the Si$_3$N$_4$ squeezing chip at low input power levels (Methods). A typical transmission spectra near the squeezing generation wavelength is shown in Fig.~\ref{fig:classical}a. Three resonances at 1559.5 nm (signal), 1560.0 nm (pump), and 1560.5 nm (idler) are selected for subsequent TMSV generation. Resonance fitting in Fig.~\ref{fig:classical}b reveals a loaded quality factor of $Q_{\rm L} \sim 0.83\times 10^6$ together with a high intrinsic quality factors $Q_{\rm i} \sim 10.1 \times 10^6$ that yield a large field escape efficiency exceeding $\eta > 91\%$, enabling the generation of strong on-chip squeezing, while the relatively high $Q_{\rm L}$ allows operation at moderate pump power levels to mitigate parasitic effects such as photorefractive noise. To assess wafer-scale uniformity, we extract the statistical distributions of $Q_{\rm i}$, $Q_{\rm L}$, and $\eta$ across resonances in the 1550–1580 nm range, as shown in Figs. 3(c–e). The most probable values for the quality factors are $Q_{\rm i} = 9.2\times10^6$ and $Q_{\rm L} = 0.9\times10^6$, demonstrating uniform low optical loss and reliable strong bus-to-resonator coupling across the wafer. Combining $Q_{\rm i}$ and $Q_{\rm L}$ infers the most probable $\eta \approx 91\%$, corresponding to a maximum achievable on-chip squeezing level as high as 10.5 dB. Notably, even the lowest $\eta \approx 75\%$ in Fig.~\ref{fig:classical}(e) promises an achievable on-chip squeezing level exceeding 6.0 dB. Moreover, the reduced material absorption and scattering at longer wavelengths are expected to yield even higher $Q_{\rm i}$'s and lower $Q_{\rm L}$'s, leading to stronger $\eta$.

We design the pump rejection filters with a relatively broad linewidth of $\sim$26 pm (corresponding to $Q_{\rm L}\sim 6\times10^4$), which reduces intracavity power buildup, mitigates thermo-optic instabilities, and increases tolerance to resonance drift. Moreover, because coupling losses dominate over intrinsic losses, the high extinction ratios (measured ranging from 30 to 40 dB) remain robust during squeezing measurement under high input powers and various wavelength range (see Supplemental Information). In addition, the edge couplers provide a reliable waveguide-to-fiber efficiency of $\sim$75\% across the wafer (see Supplemental Information), with further improvements possible through anti-reflection coatings to suppress facet reflections. Collectively, these classical characterizations validate the key requirements for the measured high squeezing across the 4-inch Si$_3$N$_4$ wafer: large escape efficiencies with high-$Q$ resonances, efficient and power-stable pump rejection, and low-loss fiber interfacing.

\subsection{On-chip squeezing-light generation: Models and characterization}
The mechanism of on-chip squeezed-light generation via non-degenerate four-wave mixing (FWM) was established in Refs. \cite{wu2020quantum,chembo2016quantum}. Following this approach, we provide a theoretical study of the squeezing performance attainable under different dispersion regimes, explicitly incorporating experimentally determined device parameters and losses. The interactions among cavity modes are governed by the quantum coupled-mode equations:
\begin{equation}
 \begin{aligned}
    \frac{d}{dt}\hat{A}_l = &-\left(\frac{\kappa}{2} + i\Delta\right)\hat{A}_l - i\frac{D_2}{2}l^2\hat{A}_l + \delta_{l,0} \sqrt{\kappa_e}A_{\rm in} \\&+ ig_0\sum_{m,n,p}\delta_{m+p,n+l}\hat{A}_m\hat{A}_n^\dagger \hat{A}_p \\&+ \sqrt{\kappa_e} \hat{V}_{e,l} + \sqrt{\kappa_i} \hat{V}_{i,l}
\end{aligned}   
\label{eq:th1}
\end{equation}
Here, $\delta_{i,j}$ is the Kronecker delta, $\hat{A}_l$ denotes the annihilation operator of the $l$th cavity mode with $l=0$ being the pump mode, $A{\rm in}$ is the input field in the bus waveguide, and $\kappa=\kappa_i+\kappa_e$ is the total linewidth with intrinsic loss $\kappa_i$ and external coupling loss $\kappa_e$. The pump detuning is represented by $\Delta$, $D_2$ is the group velocity dispersion, and the nonlinear coupling coefficient is given by $g_0 = {\hbar \omega_0^2 n_2}/{n_0^2 V_{\mathrm{eff}}}$. The operators $\hat{V}_{e,l}$ and $\hat{V}_{i,l}$ account for vacuum fluctuations introduced by external and intrinsic losses, respectively.
In the below-threshold regime, only the pump mode maintains a classical mean field, and the intracavity fields can be linearized as
\begin{equation}
   \left\{
\begin{aligned}
&  \hat{A}_0 = A_0 + \hat{a}_0(t)\\
&  \hat{A}_l = \hat{a}_l(t), l\neq 0
\end{aligned}
\right.
\label{eq:th2}
\end{equation}
where $\hat{a}_l$ denotes the quantum fluctuation of the $l$th mode. Substitution of Eq.(\ref{eq:th2}) into Eq.(\ref{eq:th1}) yields the steady-state condition for the pump:
\begin{equation}
    A_0\left(-\frac{\kappa}{2} + i\Delta\right) + \sqrt{\kappa_e}A_{\rm in} + ig_0\lvert A_0 \rvert^2 A_0 = 0
    \label{eq:th3}
\end{equation}
along with the fluctuation dynamics for the side modes:
\begin{equation}
    \frac{d}{dt}\hat{a}_l = -\left(\frac{\kappa}{2} + i\Delta\right)\hat{a}_l - i\frac{D_2}{2}l^2\hat{a}_l + 2ig_0\lvert A_0 \rvert^2\hat{a}_l + ig_0A_0^2\hat{a}_{-l}^\dagger
    \label{eq:th4}
\end{equation}
Equation~(\ref{eq:th4}) indicates that each mode $l$ is coupled exclusively with its symmetric partner $-l$, and different pairs evolve independently. The pump dynamics are governed mainly by self-phase modulation, while the side modes are driven through cross-phase modulation and FWM. Solving these coupled equations provides access to the quadrature variances of the resulting two-mode squeezed vacuum states.

\begin{figure}[bth!]
\centering
\includegraphics[width=1\textwidth]{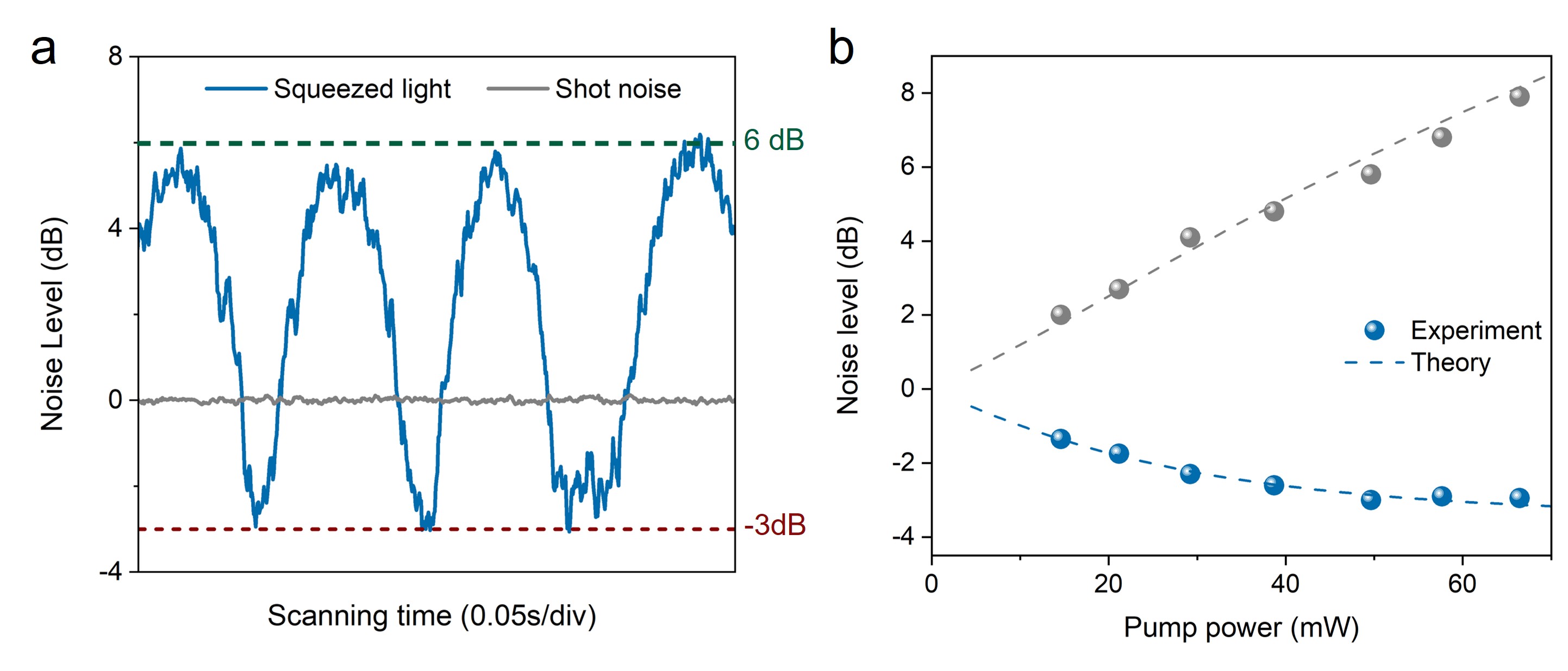}
\caption{(a) Direct measurement of the quadrature noise level at 7 MHz as the LO phase is scanned. The minimum and maximum correspond to the squeezed and anti-squeezed quadratures, showing $>$3 dB squeezing and 6.0 dB anti-squeezing relative to the shot-noise level (gray trace). (b) Measured squeezing (blue) and anti-squeezing (gray) levels as functions of on-chip pump power.  Dashed curves: theoretical model; Dots: experimental data. }\label{fig:fitting}
\end{figure}

Fig.~\ref{fig:fitting}a shows the measured quadrature noise variance of the squeezed light, normalized to this shot-noise reference. By periodically scanning the LO phase, we access all quadratures of the TMSV state. A raw squeezing level of 3.0 dB and an anti-squeezing level of 6.0 dB are directly observed at an on-chip pump power of 50 mW. Fig.~\ref{fig:fitting}b presents the measured maximum squeezing and anti-squeezing levels as a function of on-chip pump power, all remaining below the parametric oscillation threshold. As expected, both the squeezing and anti-squeezing levels increase with pump power. Beyond the optimal pump power, however, the observed squeezing slightly degrades as excess variance in the anti-squeezing quadrature mixes into the squeezing quadrature. The experimental data agree well with theoretical predictions depicted as dashed lines in Fig.~\ref{fig:fitting}b based on a model that incorporates the measured device parameters (Methods). Given the large escape efficiency of the squeezer ($\sim$91\%), the observed squeezing is primarily limited by the total collection efficiency of $\sim$60\% after generation, comprising a waveguide-to-fiber coupling efficiency of 75\%, propagation efficiency of 95\% through fiber and free-space optics, interference visibility of 98\%, and photodiode quantum efficiency of 88\%. Compared with our previous TMSV demonstration employing off-chip pump filters, this integrated design yields an improvement of more than 2 dB in measured squeezing by virtue of the ultra-low-loss integrated filters and their enabled removal of isolators. Notably, the strong quadrature squeezing is not restricted to nearest modes adjacent to the pump as the non-degenerate FWM process supports concurrent generation of broadband squeezing across multiple mode pairs known as quantum microcomb \cite {yang2021squeezed}. Such broadband squeezing modes could be accessed either by cascading EOMs and function generators to synthesize matched LOs, or by employing coherent optical soliton microcombs realized on the same 800-nm, anomalous-dispersion Si$_3$N$_4$ platform to overcome the bandwidth limitations of EO combs \cite{wu2020quantum}. 

\section{Discussion and Conclusions}

In summary, we have demonstrated wafer-scale, fully CMOS-compatible Si$_3$N$_4$ PICs for the generation of TMSV states, achieving a maximum (minimum) measured quadrature squeezing of 3.1 dB (2.9 dB) across an entire 4-inch wafer, evidencing highly uniform, robust performance. This reproducibility arises from the co-integration of ultralow-loss, strongly overcoupled microring squeezers, high-extinction pump rejection filters, low-loss edge couplers, and thermo-optic tuning on the same chip. The experimental results are in quantitative agreement with theoretical modeling based on independently extracted device parameters. The observed squeezing is currently limited by an overall off-chip detection efficiency of $\sim$60\%, pointing to a clear pathway toward further improvement through reducing extrinsic losses associated with, e.g., the fiber–chip interfaces and photodetector. In addition to the strong squeezing, the Si$_3$N$_4$ platform endows additional functionalities that are out of reach in a table-top or single integrated component platform, such as reconfigurable interferometers \cite{taballione20198}, long delay lines for time-domain cluster-state generation~\cite{asavanant2019generation,larsen2019deterministic}, and co-integration with modulators and detectors \cite{psiquantum2025manufacturable, snigirev2023ultrafast}. Such capabilities point toward a future where entire continuous-variable quantum processors can be realized on a single CMOS-compatible photonic chip, bridging the gap between laboratory-scale demonstrations and scalable quantum technologies.

\section*{Acknowledgments}
This work was supported by the National Science Foundation under Grant No.~2326780, No.~2330310, and No.~2317471 and University of Michigan.




\bibliography{myref}

\begin{thebibliography}{10}
\expandafter\ifx\csname url\endcsname\relax
  \def\url#1{\burl{#1}}\fi
\expandafter\ifx\csname urlprefix\endcsname\relax\def\urlprefix{URL }\fi
\providecommand{\bibinfo}[2]{#2}
\providecommand{\eprint}[2][]{\url{#2}}
\providecommand{\doi}[1]{\url{https://doi.org/#1}}
\bibcommenthead

\bibitem{arute2019quantum}
\bibinfo{author}{Arute, F.} \emph{et~al.}
\newblock \bibinfo{title}{Quantum supremacy using a programmable
  superconducting processor}.
\newblock \emph{\bibinfo{journal}{Nature}}
  \textbf{\bibinfo{volume}{574}}~(7779), \bibinfo{pages}{505--510}
  (\bibinfo{year}{2019}) .

\bibitem{madsen2022quantum}
\bibinfo{author}{Madsen, L.~S.} \emph{et~al.}
\newblock \bibinfo{title}{Quantum computational advantage with a programmable
  photonic processor}.
\newblock \emph{\bibinfo{journal}{Nature}}
  \textbf{\bibinfo{volume}{606}}~(7912), \bibinfo{pages}{75--81}
  (\bibinfo{year}{2022}) .

\bibitem{pirandola2018advances}
\bibinfo{author}{Pirandola, S.}, \bibinfo{author}{Bardhan, B.~R.},
  \bibinfo{author}{Gehring, T.}, \bibinfo{author}{Weedbrook, C.} \&
  \bibinfo{author}{Lloyd, S.}
\newblock \bibinfo{title}{Advances in photonic quantum sensing}.
\newblock \emph{\bibinfo{journal}{Nature Photonics}}
  \textbf{\bibinfo{volume}{12}}~(12), \bibinfo{pages}{724--733}
  (\bibinfo{year}{2018}) .

\bibitem{degen2017quantum}
\bibinfo{author}{Degen, C.~L.}, \bibinfo{author}{Reinhard, F.} \&
  \bibinfo{author}{Cappellaro, P.}
\newblock \bibinfo{title}{Quantum sensing}.
\newblock \emph{\bibinfo{journal}{Reviews of Modern Physics}}
  \textbf{\bibinfo{volume}{89}}~(3), \bibinfo{pages}{035002}
  (\bibinfo{year}{2017}) .

\bibitem{gisin2007quantum}
\bibinfo{author}{Gisin, N.} \& \bibinfo{author}{Thew, R.}
\newblock \bibinfo{title}{Quantum communication}.
\newblock \emph{\bibinfo{journal}{Nature Photonics}}
  \textbf{\bibinfo{volume}{1}}~(3), \bibinfo{pages}{165--171}
  (\bibinfo{year}{2007}) .

\bibitem{walls1983squeezed}
\bibinfo{author}{Walls, D.~F.}
\newblock \bibinfo{title}{Squeezed states of light}.
\newblock \emph{\bibinfo{journal}{Nature}}
  \textbf{\bibinfo{volume}{306}}~(5939), \bibinfo{pages}{141--146}
  (\bibinfo{year}{1983}) .

\bibitem{villar2005generation}
\bibinfo{author}{Villar, A. d.~S.}, \bibinfo{author}{Cruz, L.},
  \bibinfo{author}{Cassemiro, K.~N.}, \bibinfo{author}{Martinelli, M.} \&
  \bibinfo{author}{Nussenzveig, P.}
\newblock \bibinfo{title}{Generation of bright two-color continuous variable
  entanglement}.
\newblock \emph{\bibinfo{journal}{Physical review letters}}
  \textbf{\bibinfo{volume}{95}}~(24), \bibinfo{pages}{243603}
  (\bibinfo{year}{2005}) .

\bibitem{lawrie2019quantum}
\bibinfo{author}{Lawrie, B.~J.}, \bibinfo{author}{Lett, P.~D.},
  \bibinfo{author}{Marino, A.~M.} \& \bibinfo{author}{Pooser, R.~C.}
\newblock \bibinfo{title}{Quantum sensing with squeezed light}.
\newblock \emph{\bibinfo{journal}{Acs Photonics}}
  \textbf{\bibinfo{volume}{6}}~(6), \bibinfo{pages}{1307--1318}
  (\bibinfo{year}{2019}) .

\bibitem{o2009photonic}
\bibinfo{author}{O'brien, J.~L.}, \bibinfo{author}{Furusawa, A.} \&
  \bibinfo{author}{Vu{\v{c}}kovi{\'c}, J.}
\newblock \bibinfo{title}{Photonic quantum technologies}.
\newblock \emph{\bibinfo{journal}{Nature Photonics}}
  \textbf{\bibinfo{volume}{3}}~(12), \bibinfo{pages}{687--695}
  (\bibinfo{year}{2009}) .

\bibitem{masada2015continuous}
\bibinfo{author}{Masada, G.} \emph{et~al.}
\newblock \bibinfo{title}{Continuous-variable entanglement on a chip}.
\newblock \emph{\bibinfo{journal}{Nature Photonics}}
  \textbf{\bibinfo{volume}{9}}~(5), \bibinfo{pages}{316--319}
  (\bibinfo{year}{2015}) .

\bibitem{slusher1985observation}
\bibinfo{author}{Slusher, R.}, \bibinfo{author}{Hollberg, L.},
  \bibinfo{author}{Yurke, B.}, \bibinfo{author}{Mertz, J.} \&
  \bibinfo{author}{Valley, J.}
\newblock \bibinfo{title}{Observation of squeezed states generated by four-wave
  mixing in an optical cavity}.
\newblock \emph{\bibinfo{journal}{Physical review letters}}
  \textbf{\bibinfo{volume}{55}}~(22), \bibinfo{pages}{2409}
  (\bibinfo{year}{1985}) .

\bibitem{vahlbruch2008observation}
\bibinfo{author}{Vahlbruch, H.} \emph{et~al.}
\newblock \bibinfo{title}{Observation of squeezed light with 10-db
  quantum-noise reduction}.
\newblock \emph{\bibinfo{journal}{Physical Review Letters}}
  \textbf{\bibinfo{volume}{100}}~(3), \bibinfo{pages}{033602}
  (\bibinfo{year}{2008}) .

\bibitem{schonbeck201713}
\bibinfo{author}{Sch{\"o}nbeck, A.}, \bibinfo{author}{Thies, F.} \&
  \bibinfo{author}{Schnabel, R.}
\newblock \bibinfo{title}{13 db squeezed vacuum states at 1550 nm from 12 mw
  external pump power at 775 nm}.
\newblock \emph{\bibinfo{journal}{Optics Letters}}
  \textbf{\bibinfo{volume}{43}}~(1), \bibinfo{pages}{110--113}
  (\bibinfo{year}{2017}) .

\bibitem{vahlbruch2016detection}
\bibinfo{author}{Vahlbruch, H.}, \bibinfo{author}{Mehmet, M.},
  \bibinfo{author}{Danzmann, K.} \& \bibinfo{author}{Schnabel, R.}
\newblock \bibinfo{title}{Detection of 15 db squeezed states of light and their
  application for the absolute calibration of photoelectric quantum
  efficiency}.
\newblock \emph{\bibinfo{journal}{Physical Review Letters}}
  \textbf{\bibinfo{volume}{117}}~(11), \bibinfo{pages}{110801}
  (\bibinfo{year}{2016}) .

\bibitem{aasi2013enhanced}
\bibinfo{author}{Aasi, J.} \emph{et~al.}
\newblock \bibinfo{title}{Enhanced sensitivity of the ligo gravitational wave
  detector by using squeezed states of light}.
\newblock \emph{\bibinfo{journal}{Nature Photonics}}
  \textbf{\bibinfo{volume}{7}}~(8), \bibinfo{pages}{613--619}
  (\bibinfo{year}{2013}) .

\bibitem{ganapathy2023broadband}
\bibinfo{author}{Ganapathy, D.} \emph{et~al.}
\newblock \bibinfo{title}{Broadband quantum enhancement of the ligo detectors
  with frequency-dependent squeezing}.
\newblock \emph{\bibinfo{journal}{Physical Review X}}
  \textbf{\bibinfo{volume}{13}}~(4), \bibinfo{pages}{041021}
  (\bibinfo{year}{2023}) .

\bibitem{casacio2021quantum}
\bibinfo{author}{Casacio, C.~A.} \emph{et~al.}
\newblock \bibinfo{title}{Quantum-enhanced nonlinear microscopy}.
\newblock \emph{\bibinfo{journal}{Nature}}
  \textbf{\bibinfo{volume}{594}}~(7862), \bibinfo{pages}{201--206}
  (\bibinfo{year}{2021}) .

\bibitem{herman2025squeezed}
\bibinfo{author}{Herman, D.~I.} \emph{et~al.}
\newblock \bibinfo{title}{Squeezed dual-comb spectroscopy}.
\newblock \emph{\bibinfo{journal}{Science}}
  \textbf{\bibinfo{volume}{387}}~(6734), \bibinfo{pages}{653--658}
  (\bibinfo{year}{2025}) .

\bibitem{hariri2024entangled}
\bibinfo{author}{Hariri, A.} \emph{et~al.}
\newblock \bibinfo{title}{Entangled dual-comb spectroscopy}.
\newblock \emph{\bibinfo{journal}{arXiv preprint arXiv:2412.19800}}
  (\bibinfo{year}{2024}) .

\bibitem{asavanant2019generation}
\bibinfo{author}{Asavanant, W.} \emph{et~al.}
\newblock \bibinfo{title}{Generation of time-domain-multiplexed two-dimensional
  cluster state}.
\newblock \emph{\bibinfo{journal}{Science}}
  \textbf{\bibinfo{volume}{366}}~(6463), \bibinfo{pages}{373--376}
  (\bibinfo{year}{2019}) .

\bibitem{larsen2019deterministic}
\bibinfo{author}{Larsen, M.~V.}, \bibinfo{author}{Guo, X.},
  \bibinfo{author}{Breum, C.~R.}, \bibinfo{author}{Neergaard-Nielsen, J.~S.} \&
  \bibinfo{author}{Andersen, U.~L.}
\newblock \bibinfo{title}{Deterministic generation of a two-dimensional cluster
  state}.
\newblock \emph{\bibinfo{journal}{Science}}
  \textbf{\bibinfo{volume}{366}}~(6463), \bibinfo{pages}{369--372}
  (\bibinfo{year}{2019}) .

\bibitem{xia2020demonstration}
\bibinfo{author}{Xia, Y.} \emph{et~al.}
\newblock \bibinfo{title}{Demonstration of a reconfigurable entangled
  radio-frequency photonic sensor network}.
\newblock \emph{\bibinfo{journal}{Physical Review Letters}}
  \textbf{\bibinfo{volume}{124}}~(15), \bibinfo{pages}{150502}
  (\bibinfo{year}{2020}) .

\bibitem{guo2020distributed}
\bibinfo{author}{Guo, X.} \emph{et~al.}
\newblock \bibinfo{title}{Distributed quantum sensing in a continuous-variable
  entangled network}.
\newblock \emph{\bibinfo{journal}{Nature Physics}}
  \textbf{\bibinfo{volume}{16}}~(3), \bibinfo{pages}{281--284}
  (\bibinfo{year}{2020}) .

\bibitem{xia2023entanglement}
\bibinfo{author}{Xia, Y.} \emph{et~al.}
\newblock \bibinfo{title}{Entanglement-enhanced optomechanical sensing}.
\newblock \emph{\bibinfo{journal}{Nature Photonics}}
  \textbf{\bibinfo{volume}{17}}~(6), \bibinfo{pages}{470--477}
  (\bibinfo{year}{2023}) .

\bibitem{xia2021quantum}
\bibinfo{author}{Xia, Y.}, \bibinfo{author}{Li, W.}, \bibinfo{author}{Zhuang,
  Q.} \& \bibinfo{author}{Zhang, Z.}
\newblock \bibinfo{title}{Quantum-enhanced data classification with a
  variational entangled sensor network}.
\newblock \emph{\bibinfo{journal}{Physical Review X}}
  \textbf{\bibinfo{volume}{11}}~(2), \bibinfo{pages}{021047}
  (\bibinfo{year}{2021}) .

\bibitem{moody20222022}
\bibinfo{author}{Moody, G.} \emph{et~al.}
\newblock \bibinfo{title}{2022 roadmap on integrated quantum photonics}.
\newblock \emph{\bibinfo{journal}{Journal of Physics: Photonics}}
  \textbf{\bibinfo{volume}{4}}~(1), \bibinfo{pages}{012501}
  (\bibinfo{year}{2022}) .

\bibitem{wang2025large}
\bibinfo{author}{Wang, Z.} \emph{et~al.}
\newblock \bibinfo{title}{Large-scale cluster quantum microcombs}.
\newblock \emph{\bibinfo{journal}{Light: Science \& Applications}}
  \textbf{\bibinfo{volume}{14}}~(1), \bibinfo{pages}{164}
  (\bibinfo{year}{2025}) .

\bibitem{jia2025continuous}
\bibinfo{author}{Jia, X.} \emph{et~al.}
\newblock \bibinfo{title}{Continuous-variable multipartite entanglement in an
  integrated microcomb}.
\newblock \emph{\bibinfo{journal}{Nature}} \bibinfo{pages}{1--8}
  (\bibinfo{year}{2025}) .

\bibitem{tasker2021silicon}
\bibinfo{author}{Tasker, J.~F.} \emph{et~al.}
\newblock \bibinfo{title}{Silicon photonics interfaced with integrated
  electronics for 9 ghz measurement of squeezed light}.
\newblock \emph{\bibinfo{journal}{Nature Photonics}}
  \textbf{\bibinfo{volume}{15}}~(1), \bibinfo{pages}{11--15}
  (\bibinfo{year}{2021}) .

\bibitem{gurses2025chip}
\bibinfo{author}{Gurses, V.} \emph{et~al.}
\newblock \bibinfo{title}{An on-chip phased array for non-classical light}.
\newblock \emph{\bibinfo{journal}{Nature Communications}}
  \textbf{\bibinfo{volume}{16}}~(1), \bibinfo{pages}{6849}
  (\bibinfo{year}{2025}) .

\bibitem{van1999unconditional}
\bibinfo{author}{Van~Loock, P.} \& \bibinfo{author}{Braunstein, S.~L.}
\newblock \bibinfo{title}{Unconditional teleportation of continuous-variable
  entanglement}.
\newblock \emph{\bibinfo{journal}{Physical Review A}}
  \textbf{\bibinfo{volume}{61}}~(1), \bibinfo{pages}{010302}
  (\bibinfo{year}{1999}) .

\bibitem{fukui2018high}
\bibinfo{author}{Fukui, K.}, \bibinfo{author}{Tomita, A.},
  \bibinfo{author}{Okamoto, A.} \& \bibinfo{author}{Fujii, K.}
\newblock \bibinfo{title}{High-threshold fault-tolerant quantum computation
  with analog quantum error correction}.
\newblock \emph{\bibinfo{journal}{Physical Review X}}
  \textbf{\bibinfo{volume}{8}}~(2), \bibinfo{pages}{021054}
  (\bibinfo{year}{2018}) .

\bibitem{fukui2023high}
\bibinfo{author}{Fukui, K.}
\newblock \bibinfo{title}{High-threshold fault-tolerant quantum computation
  with the gottesman-kitaev-preskill qubit under noise in an optical setup}.
\newblock \emph{\bibinfo{journal}{Physical Review A}}
  \textbf{\bibinfo{volume}{107}}~(5), \bibinfo{pages}{052414}
  (\bibinfo{year}{2023}) .

\bibitem{chen2022ultra}
\bibinfo{author}{Chen, P.-K.}, \bibinfo{author}{Briggs, I.},
  \bibinfo{author}{Hou, S.} \& \bibinfo{author}{Fan, L.}
\newblock \bibinfo{title}{Ultra-broadband quadrature squeezing with thin-film
  lithium niobate nanophotonics}.
\newblock \emph{\bibinfo{journal}{Optics Letters}}
  \textbf{\bibinfo{volume}{47}}~(6), \bibinfo{pages}{1506--1509}
  (\bibinfo{year}{2022}) .

\bibitem{park2024single}
\bibinfo{author}{Park, T.} \emph{et~al.}
\newblock \bibinfo{title}{Single-mode squeezed-light generation and tomography
  with an integrated optical parametric oscillator}.
\newblock \emph{\bibinfo{journal}{Science Advances}}
  \textbf{\bibinfo{volume}{10}}~(11), \bibinfo{pages}{eadl1814}
  (\bibinfo{year}{2024}) .

\bibitem{arge2024demonstration}
\bibinfo{author}{Arge, T.~N.} \emph{et~al.}
\newblock \bibinfo{title}{Demonstration of a squeezed light source on thin-film
  lithium niobate with modal phase matching}.
\newblock \emph{\bibinfo{journal}{arXiv preprint arXiv:2406.16516}}
  (\bibinfo{year}{2024}) .

\bibitem{shi2025squeezed}
\bibinfo{author}{Shi, X.} \emph{et~al.}
\newblock \bibinfo{title}{Squeezed light generation in periodically poled
  thin-film lithium niobate waveguides}.
\newblock \emph{\bibinfo{journal}{arXiv preprint arXiv:2508.08599}}
  (\bibinfo{year}{2025}) .

\bibitem{dutt2015chip}
\bibinfo{author}{Dutt, A.} \emph{et~al.}
\newblock \bibinfo{title}{On-chip optical squeezing}.
\newblock \emph{\bibinfo{journal}{Physical Review Applied}}
  \textbf{\bibinfo{volume}{3}}~(4), \bibinfo{pages}{044005}
  (\bibinfo{year}{2015}) .

\bibitem{vaidya2020broadband}
\bibinfo{author}{Vaidya, V.~D.} \emph{et~al.}
\newblock \bibinfo{title}{Broadband quadrature-squeezed vacuum and nonclassical
  photon number correlations from a nanophotonic device}.
\newblock \emph{\bibinfo{journal}{Science Advances}}
  \textbf{\bibinfo{volume}{6}}~(39), \bibinfo{pages}{eaba9186}
  (\bibinfo{year}{2020}) .

\bibitem{zhao2020near}
\bibinfo{author}{Zhao, Y.} \emph{et~al.}
\newblock \bibinfo{title}{Near-degenerate quadrature-squeezed vacuum generation
  on a silicon-nitride chip}.
\newblock \emph{\bibinfo{journal}{Physical Review Letters}}
  \textbf{\bibinfo{volume}{124}}~(19), \bibinfo{pages}{193601}
  (\bibinfo{year}{2020}) .

\bibitem{zhang2021squeezed}
\bibinfo{author}{Zhang, Y.} \emph{et~al.}
\newblock \bibinfo{title}{Squeezed light from a nanophotonic molecule}.
\newblock \emph{\bibinfo{journal}{Nature Communications}}
  \textbf{\bibinfo{volume}{12}}~(1), \bibinfo{pages}{2233}
  (\bibinfo{year}{2021}) .

\bibitem{shen2025strong}
\bibinfo{author}{Shen, Y.} \emph{et~al.}
\newblock \bibinfo{title}{Strong nanophotonic quantum squeezing exceeding 3.5
  db in a foundry-compatible kerr microresonator}.
\newblock \emph{\bibinfo{journal}{Optica}} \textbf{\bibinfo{volume}{12}}~(3),
  \bibinfo{pages}{302--308} (\bibinfo{year}{2025}) .

\bibitem{ulanov2025quadrature}
\bibinfo{author}{Ulanov, A.~E.}, \bibinfo{author}{Ruhnke, B.},
  \bibinfo{author}{Wildi, T.} \& \bibinfo{author}{Herr, T.}
\newblock \bibinfo{title}{Quadrature squeezing in a nanophotonic
  microresonator}.
\newblock \emph{\bibinfo{journal}{arXiv preprint arXiv:2502.17337}}
  (\bibinfo{year}{2025}) .

\bibitem{yang2021squeezed}
\bibinfo{author}{Yang, Z.} \emph{et~al.}
\newblock \bibinfo{title}{A squeezed quantum microcomb on a chip}.
\newblock \emph{\bibinfo{journal}{Nature Communications}}
  \textbf{\bibinfo{volume}{12}}~(1), \bibinfo{pages}{4781}
  (\bibinfo{year}{2021}) .

\bibitem{xiang2022silicon}
\bibinfo{author}{Xiang, C.}, \bibinfo{author}{Jin, W.} \&
  \bibinfo{author}{Bowers, J.~E.}
\newblock \bibinfo{title}{Silicon nitride passive and active photonic
  integrated circuits: trends and prospects}.
\newblock \emph{\bibinfo{journal}{Photonics research}}
  \textbf{\bibinfo{volume}{10}}~(6), \bibinfo{pages}{A82--A96}
  (\bibinfo{year}{2022}) .

\bibitem{shekhar2024roadmapping}
\bibinfo{author}{Shekhar, S.} \emph{et~al.}
\newblock \bibinfo{title}{Roadmapping the next generation of silicon
  photonics}.
\newblock \emph{\bibinfo{journal}{Nature Communications}}
  \textbf{\bibinfo{volume}{15}}~(1), \bibinfo{pages}{751}
  (\bibinfo{year}{2024}) .

\bibitem{grosshans2001quantum}
\bibinfo{author}{Grosshans, F.} \& \bibinfo{author}{Grangier, P.}
\newblock \bibinfo{title}{Quantum cloning and teleportation criteria for
  continuous quantum variables}.
\newblock \emph{\bibinfo{journal}{Physical Review A}}
  \textbf{\bibinfo{volume}{64}}~(1), \bibinfo{pages}{010301}
  (\bibinfo{year}{2001}) .

\bibitem{wu2020quantum}
\bibinfo{author}{Wu, B.-H.}, \bibinfo{author}{Alexander, R.~N.},
  \bibinfo{author}{Liu, S.} \& \bibinfo{author}{Zhang, Z.}
\newblock \bibinfo{title}{Quantum computing with multidimensional
  continuous-variable cluster states in a scalable photonic platform}.
\newblock \emph{\bibinfo{journal}{Physical Review Research}}
  \textbf{\bibinfo{volume}{2}}~(2), \bibinfo{pages}{023138}
  (\bibinfo{year}{2020}) .

\bibitem{chembo2016quantum}
\bibinfo{author}{Chembo, Y.~K.}
\newblock \bibinfo{title}{Quantum dynamics of kerr optical frequency combs
  below and above threshold: Spontaneous four-wave mixing, entanglement, and
  squeezed states of light}.
\newblock \emph{\bibinfo{journal}{Physical Review A}}
  \textbf{\bibinfo{volume}{93}}~(3), \bibinfo{pages}{033820}
  (\bibinfo{year}{2016}) .

\bibitem{taballione20198}
\bibinfo{author}{Taballione, C.} \emph{et~al.}
\newblock \bibinfo{title}{8$\times$ 8 reconfigurable quantum photonic processor
  based on silicon nitride waveguides}.
\newblock \emph{\bibinfo{journal}{Optics Express}}
  \textbf{\bibinfo{volume}{27}}~(19), \bibinfo{pages}{26842--26857}
  (\bibinfo{year}{2019}) .

\bibitem{psiquantum2025manufacturable}
\bibinfo{author}{PsiQuantum}.
\newblock \bibinfo{title}{A manufacturable platform for photonic quantum
  computing}.
\newblock \emph{\bibinfo{journal}{Nature}} \textbf{\bibinfo{volume}{641}},
  \bibinfo{pages}{876–883} (\bibinfo{year}{2025}) .

\bibitem{snigirev2023ultrafast}
\bibinfo{author}{Snigirev, V.} \emph{et~al.}
\newblock \bibinfo{title}{Ultrafast tunable lasers using lithium niobate
  integrated photonics}.
\newblock \emph{\bibinfo{journal}{Nature}}
  \textbf{\bibinfo{volume}{615}}~(7952), \bibinfo{pages}{411--417}
  (\bibinfo{year}{2023}) .

\bibitem{liu2025fabrication}
\bibinfo{author}{Liu, S.}, \bibinfo{author}{Zhang, Y.},
  \bibinfo{author}{Hariri, A.}, \bibinfo{author}{Al-Hallak, A.-R.} \&
  \bibinfo{author}{Zhang, Z.}
\newblock \bibinfo{title}{Fabrication of ultra-low-loss, dispersion-engineered
  silicon nitride photonic integrated circuits via silicon hardmask etching}.
\newblock \emph{\bibinfo{journal}{ACS Photonics}}
  \textbf{\bibinfo{volume}{12}}~(2), \bibinfo{pages}{1039--1046}
  (\bibinfo{year}{2025}) .

\end{thebibliography}

\section*{Methods}\label{sec11}

\bmhead{Device fabrication process}
The complete wafer-scale fabrication of the Si$_3$N$_4$ squeezed-light chips \cite{liu2025fabrication} is performed using an amorphous silicon (a-Si) hardmask dry etching process in the University of Michigan's Lurie Nanofabrication Facility. A detailed fabrication flow is shown in Supplementary Information. The fabrication begins with the growth of a 3000-nm-thick wet thermal SiO$_2$ layer on a 4-inch bare Si substrate, followed by the deposition of a 380-nm low-pressure chemical vapor deposition (LPCVD) Si$_3$N$_4$ film. Crack-isolation trenches are defined along the wafer boundary using photolithography and inductively coupled plasma reactive ion etching (ICP-RIE), etching through both the Si$_3$N$_4$ and the underlying SiO$_2$ layer. After photoresist removal and thorough wafer cleaning, a second 420-nm LPCVD Si$_3$N$_4$ layer is deposited, resulting in a final Si$_3$N$_4$ thickness of approximately 800 nm. The combined bilayer structure is dispersion-engineered to exhibit anomalous group velocity dispersion. On-chip structures are defined in eight die regions across the wafer using electron-beam lithography (JEOL 6300) with maN-2405 resist. The patterns are transferred into the a-Si hardmask layer via ICP-RIE (LAM 9400) using a HBr/He chemistry. The underlying 800-nm Si$_3$N$_4$ film is then fully etched using a separate ICP-RIE system (STS APS DGRIE) with a gas mixture of C$_4$F$_8$, CF$_4$, and He. The residual a-Si hardmask is removed using isotropic XeF$_2$ etching (Xactix). After completely etching the backside Si$_3$N$_4$ layer, the wafer undergoes Piranha and standard RCA cleaning, followed by high-temperature annealing in N$_2$ at 1100$^{\circ}$C for 6 hours. To complete the waveguide cladding, an 800-nm LPCVD TEOS oxide layer is first deposited, followed by a 2200-nm plasma-enhanced chemical vapor deposition (PECVD) SiO$_2$ layer and another 6-hour N$_2$ annealing at 1100$^{\circ}$C. Microheaters are fabricated by depositing 5-nm Ti and 250-nm Pt films onto the SiO$_2$ cladding surface using UV stepper photolithography (GCA AS200) alignment with LOR/S1813 resist, electron-beam evaporation, and a regular lift-off process with Remover PG and IPA. To form the edge couplers, taper structures are exposed by UV stepper photolithography with AZ 12-XT, followed by full ICP-RIE etching of both the PECVD SiO$_2$ cladding and the buried SiO$_2$ layer. For chip release, the 250-$\mu$m-thick Si substrate is ICP-RIE etched using the Bosch deep silicon etching process (STS Pegasus) with alternating cycles of SF$_6$, C$_4$F$_8$, and O$_2$ gases. The wafer is then diced into individual chips for classical characterization. Finally, each chip is wire-bonded (MPP iBond 5000) to a custom PCB circuit to enable thermal tuning of the on-chip microheater elements with external multi-channel current sources for squeezing measurement.

\bmhead{Classical characterization} 
A tunable CW laser (Santec TSL-770) is passed through a fiber polarization controller and coupled into the bus waveguide of the Si$_3$N$_4$ squeezing chip using a lensed fiber. The laser wavelength is swept at 20 nm/s and the transmitted light through the squeezer and integrated filters is collected by another lensed fiber and detected with a low-noise photodetector (NewFocus 1811). The on-chip optical power is maintained below 20 $\mu$W to avoid thermal effects. The transmission spectra are recorded on an oscilloscope and analyzed using custom Python scripts to extract the coupling and intrinsic quality factors ($Q_\mathrm{c}$ and $Q_\mathrm{i}$) of the target resonances via Lorentzian fitting. 

\bmhead{Quantum characterization} 
A low-noise tunable CW laser (TLB-6700) is amplified by an erbium-doped fiber amplifier (EDFA) and subsequently filtered by a free-space, narrow-linewidth ($\sim$ 2 MHz) mode cleaner to suppress amplified spontaneous emission (ASE) noise. The output light is then split into two paths: one serves as pump for the Si$_3$N$_4$ photonic chip to generate the TMSV states, while the other is used for preparing the LO for balanced homodyne detection of the generated squeezed light. 

In the Si$_3$N$_4$ squeezing chip path, the light passes through two polarization controllers to excite the TE$_{00}$ bus waveguide mode. To align the target squeezing resonance with the pump laser, the microring is first initialized at an elevated temperature, shifting its resonance to longer wavelengths. By gradually reducing the microheater temperature, the cavity resonance is blue-shifted to align with the pump light for efficient squeezing generation. Below the parametric oscillation threshold, TMSV states are generated and escape the squeezer together with residual pump light. The unwanted pump is filtered by the two filters aligned with the pump wavelength. We monitor the power at the two drop ports for actively feedback control of the microheaters. Each filter achieves 30-40 dB of filtering efficiency, rendering a single filter adequate to reduce the residual pump power below 50 $\mu$W in most cases. The insertion loss is negligible owing to the straight bus waveguide. The deep thermal isolation trenches effectively eliminate thermal crosstalk, ensuring stable and reproducible operation of the filters without affecting the squeezer. The squeezed light is subsequently collected by a lensed fiber and launched into a free-space path for homodyne measurement. 

In the LO path, the pump light is sent through a fiber EOM (EOspace) driven by a radio frequency (RF) function generator. A modulation frequency close to $f_\mathrm{m} \sim 19.77$ GHz is chosen such that the generated third-order sideband aligns with the free spectral range of the squeezer ($3f_\mathrm{m} \sim \mathrm{FSR}$). The desired LO sideband tooth is filtered using a programmable waveshaper (Finisar) with a specified extinction ratio exceeding 60 dB. During the experiment, $f_\mathrm{m}$ is fine tuned to achieve the maximum amount of squeezing. The LO, with optical power of approximately 5 mW, is then combined with the two-mode squeezed light on a 50:50 beam splitter, and the outputs are detected with a pair photodiodes operating in a balanced configuration. The difference photocurrent is then analyzed on an electrical spectrum analyzer (ESA) at a central frequency of 7 MHz, resolution bandwidth (RBW) of 300 kHz, and video bandwidth (VBW) of 470 Hz. The common-mode rejection ratio of the balanced detectors is measured to be $\sim$ 40 dB, allowing the homodyne detection to operation in the shot-noise-limited regime calibrated by injecting only the LO into the balanced photodiodes. 

\bmhead{Model for on-chip squeezed-light generation} 
To formulate a quantum-optical model of squeezed-light generation, we start with a semiclassical picture prior to canonical quantization. The classical coupled-mode relation is
\begin{equation}
    \frac{dA_l}{dt} = -\left(\frac{\kappa}{2} + i\Delta\right)A_l - i\frac{D_2}{2}l^2A_l + \delta(l,0)\sqrt{\kappa_e}A_{\rm in} + ig_0\sum_{m,n,p}\delta_{m+p,n+l}A_mA_n^*A_p.
    \label{eq:th5}
\end{equation}
Here, $A_l$ is the amplitude of the $l$-th mode with respect to the pump, normalized so that $\vert A_l \vert^2$ represents photon flux. The resonance linewidth $\kappa$ captures both the intrinsic loss $\kappa_i$ and coupling loss $\kappa_e$. The pump detuning is $\Delta$, the dispersion is quantified by $D_2$, and the nonlinear coefficient $g_0$ gives the per-photon frequency shift from four-wave mixing. We assume perfect overlap of the azimuthal modes. Eq.~(\ref{eq:th5}) therefore models the dynamics and coupling of cavity modes, laying the groundwork for analyzing two-mode squeezing. Following Ref.~\cite{chembo2016quantum}, canonical quantization is carried out by promoting $A$ and $A^*$ to operators $\hat A$ and $\hat A^\dagger$, in conjuction with adding vacuum noise operators $\hat V$ for each loss channel to ensure the commutation relation persists, giving rise to the quantum version of the coupled-mode equations:
\begin{equation}
 \begin{aligned}
    \frac{d}{dt}\hat{A}_l = &-\left(\frac{\kappa}{2} + i\Delta\right)\hat{A}_l - i\frac{D_2}{2}l^2\hat{A}_l + \delta_{l,0}\sqrt{\kappa_e}A_{\rm in} \\&+ ig_0\sum_{m,n,p}\delta_{m+p,n+l}\hat{A}_m\hat{A}_n^\dagger \hat{A}_p \\&+ \sqrt{\kappa_e} \hat{V}_{e,l} + \sqrt{\kappa_i} \hat{V}_{i,l}.
\end{aligned}   
\label{eq:th6}
\end{equation}
Operating below the optical parametric oscillation threshold, only the pumped mode acquires a mean field, while all other modes only exhibit quantum fluctuations. Thus, the operators can be linearized as
\begin{equation}
   \left\{
\begin{aligned}
&  \hat{A_0} = A_0 + \hat{a}_0(t)\\
&  \hat{A_l} = \hat{a}_l(t), l\neq 0.
\end{aligned}
\right.
\label{eq:th7}
\end{equation}
Substituting Eq.(\ref{eq:th7}) into Eq.(\ref{eq:th6}) gives the steady-state condition
\begin{equation}
    A_0\left(-\frac{\kappa}{2} + i\Delta\right) + \sqrt{\kappa_e}A_{\rm in} + ig_0\vert A_0\vert^2A_0 = 0,
    \label{eq:th8}
\end{equation}
which determines the classical amplitude $A_0$. The quantum fluctuation dynamics of other modes follow as
\begin{equation}
    \frac{d}{dt}\hat{a}_l = -\left(\frac{\kappa}{2} + i\Delta\right)\hat{a}_l - i\frac{D_2}{2}l^2\hat{a}_l + 2ig_0\vert A_0 \vert ^2\hat{a}_l + ig_0A_0^2\hat{a}_{-l}^\dagger.
    \label{eq:th9}
\end{equation}
These equations can equivalently be obtained from the Hamiltonian in the Heisenberg picture~\cite{chembo2016quantum}. Nonlinear terms show that the pump is only governed by self-phase modulation, while sideband fluctuations arise through both cross-phase modulation and four-wave mixing. Importantly, each pair of $l$ and $-l$ modes is coupled but independent of other pairs.

We now transform into the frequency domain. The output operator for mode $l$ reads
\begin{equation}
    \hat{\tilde{A}}_{\mathrm{out},l}(\omega) = \sqrt{\kappa_e} \hat{\tilde{A}}_{l}(\omega) - \hat{\tilde{V}}_{e,l}(\omega),
\end{equation}
with the vacuum operators obeying the following correlations:
\begin{equation}
\begin{aligned}
    &\langle\hat{\tilde{V}}_{t,l}(\omega)\hat{\tilde{V}}_{t',l'}^\dagger(\omega')\rangle = \delta_{t,t'}\delta_{l,l'}\delta(\omega-\omega')\\&\langle\hat{\tilde{V}}_{t,l}^\dagger(\omega)\hat{\tilde{V}}_{t',l'}(\omega')\rangle = 0\\&\langle\hat{\tilde{V}}_{t,l}(\omega)\hat{\tilde{V}}_{t',l'}(\omega')\rangle = \langle\hat{\tilde{V}}_{t,l}^\dagger(\omega)\hat{\tilde{V}}_{t',l'}^\dagger(\omega')\rangle = 0; t = i,e.
\end{aligned}
\end{equation}
Solving Eq.~(\ref{eq:th9}) allows one to extract the fluctuation spectra and thereby the squeezing spectrum for the $l$ and $-l$ modes.

To account for detection, all losses are combined into an overall quantum efficiency $\eta_{\rm total}$, so the measured operator is
\begin{equation}
    \hat{\tilde{A}}_{\mathrm{meas},l} = \sqrt{\eta_{\mathrm{total}}} ~\hat{\tilde A}_{\mathrm{out},l} + \sqrt{1-\eta_{\mathrm{total}}}~\hat{\tilde V}_{\eta}
\end{equation}
where $\hat{\tilde{A}}_{\mathrm{meas},l}$ is the detected operator, and $\hat V_\eta$ represents additional vacuum noise.

Finally, in balanced homodyne detection, the LO carries components at $\pm l$. Introducing rotated quadratures for LO with phase $\theta$:
\begin{equation}
\begin{aligned}
    \hat{\tilde q}_l(\omega ; \theta ) = \hat{\tilde q}_l(\omega)\cos{\theta} - \hat{\tilde p}_l(\omega)\sin{\theta}, \\
    \hat{\tilde p}_l(\omega ; \theta ) = \hat{\tilde q}_l(\omega)\sin{\theta} + \hat{\tilde p}_l(\omega)\cos{\theta},
\end{aligned}   
\end{equation}
with definitions
\begin{equation}
    \begin{aligned}
        \hat{\tilde q}_l(\omega) &= \frac{1}{2}(\hat{\tilde{A}}_{\mathrm{meas},l}(\omega) + \hat{\tilde{A}}^\dagger_{\mathrm{meas},l}(-\omega)) \\
        \hat{\tilde p}_l(\omega) &= \frac{1}{2i}(\hat{\tilde{A}}_{\mathrm{meas},l}(\omega) - \hat{\tilde{A}}^\dagger_{\mathrm{meas},l}(-\omega)),
    \end{aligned}
\end{equation}
the detector output is found proportional to
\begin{equation}
    \hat{\tilde X}_{\theta}(\omega) = [\hat{\tilde q}_l(\omega ; \theta ) + \hat{\tilde q}_{-l}(\omega ; \theta )]\cos{\theta} + [\hat{\tilde p}_l(\omega ; \theta ) + \hat{\tilde p}_{-l}(\omega ; \theta )]\sin{\theta},
\end{equation}
providing direct access to the squeezing and anti-squeezing levels.



\backmatter

\noindent

\bigskip

\begin{appendices}

\end{appendices}

\end{document}